\documentclass[graybox]{svmult}

\usepackage{mathptmx}       
\usepackage{helvet}         
\usepackage{courier}        
\usepackage{type1cm}        
\usepackage{makeidx}         
\usepackage{graphicx}        
\usepackage{multicol}        
\usepackage[bottom]{footmisc}

\makeindex             

\begin{document}

\title*{The influence of cosmic rays in the circumnuclear molecular gas of NGC\,1068}
\author{Rebeca Aladro, Serena Viti, Estelle Bayet, Denise Riquelme}
\institute{Rebeca Aladro \& Serena Viti \at Dept. of Physics \&  Astronomy, UCL, Gower Street, London WC1E 6BT, UK, \email{r.aladro@ucl.ac.uk, sv@star.ucl.ac.uk}
\and Estelle Bayet \at Sub-Department of Astrophysics, University of Oxford, Denys Wilkinson Building, Keble Road, Oxford OX1 3RH, UK, \email{Estelle.Bayet@astro.ox.ac.uk}
\and Denise Riquelme \at Instituto de Radioastronom\'ia Milim\'etrica, Avda. Divina Pastora, 7, Local 20, E-18012 Granada, Spain, \email riquelme@iram.es}
%
%
\maketitle

\abstract*{We surveyed the circumnuclear disk of the Seyfert galaxy NGC\,1068 between the frequencies 86.2\,GHz and 115.6\,GHz, and identified 17 different molecules. Using a time and depth dependent chemical model we reproduced the observational results, and  show that the column densities of most of the species are better reproduced if the cold gas is heavily pervaded by a high cosmic ray ionization rate of about  1000 times that of the Milky Way. We discuss how molecules in the NGC\,1068 nucleus may be influenced by this external radiation, as well as by UV radiation fields.}

\abstract{We surveyed the circumnuclear disk of the Seyfert galaxy NGC\,1068 between the frequencies 86.2\,GHz and 115.6\,GHz, and identified 17 different molecules. Using a time and depth dependent chemical model we reproduced the observational results, and  show that the column densities of most of the species are better reproduced if the cold gas is heavily pervaded by a high cosmic ray ionization rate of about  1000 times that of the Milky Way. We discuss how molecules in the NGC\,1068 nucleus may be influenced by this external radiation, as well as by UV radiation fields.}

\section{Introduction}
\label{Intro}

The NGC\,1068 proximity (D=14.4\,Mpc, \cite{Bland-Hawthorn97}) and luminosity ($L_{\rm IR}=3\times10^{11} L_\odot$, \cite{Telesco80}) make this galaxy the best studied object in the local universe hosting an active galactic nucleus (AGN). Besides, NGC\,1068 shows an extremely rich molecular complexity, as shown by many studies conducted in the millimeter (mm) and sub-mm ranges (e.g. \cite{Kamenetzky11,Costagliola11,Nakajima11}). Molecular emission is one of the best tools to study the physical properties of the interstellar medium in galaxy nuclei, and in particular in AGN where the gas is heavily obscured by large amounts of dust that absorb the emitted radiation in other wavelengths. The gas in the circumnuclear disk (CND) of NGC\,1068 seems to be heavily pervaded by X-ray radiation originated in the nuclear accretion disk, forming a giant X-ray dominated region (XDR) which leaves particular imprints in the chemical composition of the ISM, such as enhanced SiO and CN abundances \cite{Usero04,Burillo10}. To understand the role of X-rays in this type of galaxies, we conducted an unbiased molecular line survey towards the CND of NGC\,1068, and modelled its molecular emission considering different combinations of far UV (FUV) radiation fields and cosmic ray ionization rates.

\section{Observations, data reduction and analysis}
\label{obs}

We observed the CND of NGC\,1068 ($\alpha_{2000}=$\,02:42:40.9, $\delta_{2000}$=\,-00:00:46.0) with the IRAM 30-m telescope\footnote{IRAM is supported by INSU/CNRS (France), MPG (Germany) and IGN (Spain).} (Pico Veleta Observatory, Spain) between October 2009 and July 2010. Using the band E0 of the EMIR receiver and the WILMA autocorrelator, we covered the frequencies between 86.2\,GHz and 115.6\,GHz, for which the telescope beam size ranged from 21$''$ to 29$''$, and the channel width spacing was $7-9$\,km\,s$^{-1}$. The data was calibrated using the standard dual method. The observations were done wobbling the secondary mirror with a switching frequency of 0.5\,Hz and a beam throw of 220$''$ in azimuth. We checked the pointing accuracy every hour towards nearby bright continuum sources. The pointing corrections were always better than 4$''$. The focus was also checked at the beginning of each run and during sunsets. 

The observed spectra were converted from antenna temperatures ($T_{\rm A}^*$) to main beam temperatures ($T_{\rm MB}$) using the relation $T_{\rm MB}=(F_{\rm eff}/B_{\rm eff})\,T_{\rm A}^*$, where $F_{\rm eff}$ is the forward efficiency of the telescope, whose values were between 0.94 and 0.95, and $B_{\rm eff}$ is the main beam efficiency, ranging from 0.77 to 0.81. Linear baselines were subtracted in all cases. The rms achieved is $\le 2$\,mK across the whole survey. The data were also corrected by beam dilution effects as $T_{\rm B}=[(\theta^2_{\rm s}\,+\,\theta^2_{\,\rm b})\,/\,\theta^2_{\,\rm s}]\,T_{\rm MB}$, where $T_{\rm B}$ is the source averaged brightness temperature, $\theta_{\,\rm s}$ is the source size and $\theta_{\,\rm b}$ is the beam size. Based on NGC\,1068 interferometric observations of $^{12}$CO, HCN and $^{13}$CO (e.g. \cite{Helfer95,Schinnerer00}) we have assumed an average source size of 4$''$ for all the species detected in this survey. 

Gaussian profiles were fitted to all the detected lines. 
The reduction of the spectra and Gaussian profile fitting were done using the CLASS \footnote{CLASS $http://www.iram.fr/IRAMFR/GILDAS$} and MASSA\footnote{MASSA $http://damir.iem.csic.es/mediawiki-1.12.0/index.php/MASSA\_User's\_Manual$} software packages. 
Figures showing the spectra, the resulting parameters from the Gaussian fittings, and details about each detected molecule are shown in \cite{Aladro12}.

Assuming local thermodynamic equilibrium (LTE) and optically thin emission for all the detected species, we made Boltzmann diagrams to roughly estimate the column densities of the molecules. However some species, such as CO and HCN, are certainly affected by optically thickness. The values of the column densities for each species, as well as the impact of opacity effects in our calculations, are also shown in \cite{Aladro12}.

\section{Modelling of the data}
\label{modelling}

 \begin{table}
\caption{Parameters that characterise our different UCL\_CHEM models}
\label{UCLCHEM}
\begin{tabular}[!h]{lcccccc} 
\hline
Parameter	& Model a	 &	Model b	& Model c & Model d	\\
\noalign{\smallskip}\svhline\noalign{\smallskip}

Temperature (Phase II)$^\dagger$	& 200/100\,K & 300/100\,K & 250/300\,K & 350/300\,K &\\
Visual extinction	& 2 \& 10\,mag &  2 \& 10\,mag & 2 \& 10\,mag & 2 \& 10\,mag\\
External UV radiation intensity	& 1 Habing & 1000 Habing & 1 Habing & 1000 Habing\\
Cosmic-ray ionization rate ($\zeta$)&	1.3$\times$10$^{-17}$\,s$^{-1}$ & 1.3$\times$10$^{-17}$\,s$^{-1}$ & 1.3$\times$10$^{-15}$\,s$^{-1}$ & 1.3$\times$10$^{-15}$\,s$^{-1}$ \\
\noalign{\smallskip}\hline\noalign{\smallskip}
 \end{tabular}
\\$^\dagger$ First value corresponds to A$_v=2$\,mag and second to A$_v=10$\,mag.
\end{table}

To model the observations we used the time and depth dependent chemical model UCL\_CHEM \cite{Viti99,Viti04}, which was run in two separate phases. During Phase I, we first simulated the formation of a dark cloud by collapsing an atomic diffuse gas from a density of 100 cm$^{-3}$ to a final density of $10^5$cm$^{-3}$. During the collapse atoms and molecules freeze onto the dust grains forming icy mantles. Both gas and surface chemistries are self-consistently computed. Once the gas is in equilibrium again, UCL\_CHEM compiles in Phase II the chemical evolution of the gas and the dust after a burst of star formation has occurred, so, while the temperature during the first phase of the modelling was kept to 10K, during the second phase
it is increased to 100-350 K  and the icy mantles are evaporated. The chemical evolution of the gas is then followed for 10$^7$ years. In both phases of the UCL\_CHEM model the chemical
network is based on more than 2345 chemical reactions involving 205
species of which 51 are surface species. The gas phase reactions were adopted from the UMIST data base \cite{Millar97, LeTeuff00, Woodall07}. The surface reactions included in this model are assumed to be mainly hydrogenation reactions, 
allowing chemical saturation where this is
possible.

To sample the likely co-existing conditions that molecules are experiencing in the
nucleus of NGC 1068, we run four models (named a, b, c, and d), which  allowed us to investigate the response of the chemistry to the changes in FUV radiation fields and  cosmic-rays
ionization rate, $\zeta$ (which can be used to simulate XDR-like environments). The values of these two parameters are listed in Table~\ref{UCLCHEM}. On the other hand, in order to reproduce the physical conditions of the CND of NGC\,1068, we used a final density of n$\rm _H$=$10^5$\,cm$^{-2}$ and a metallicity value of z=1.056\,z$_\odot$, based on previous studies of the galaxy (e.g. \cite{Zaritsky94,Tacconi94,Kamenetzky11}). The initial elemental
abundances (C/H, O/H, N/H, S/H, He/H, Mg/H and Si/H) used in our study
are those corresponding to extragalactic environments as described in
\cite{Bayet08}. Other parameters not mentioned (e.g. grain size)
have been kept to their standard (Milky Way) values.

Using the results of the models, we calculated the column densities of
the seventeen species detected in our survey at a representative time of $10^5$\,yr (but note that chemistry has not necessarily reached steady 
state by then), using the formula: \\
\begin{equation}
N_{mol} = X_{mol} \times A_V \times 1.6\times10^{21}
\end{equation}

where $X_{\rm mol}$ is the fractional abundance of the molecule, $A_{\rm V}$ is the visual extinction, and  $1.6\times10^{21}$ is the column density of hydrogen at a visual extinction of one magnitude.
Note that the formulation above simply give an `on the spot' approximation of the column density. 
We set $N_{\rm mol}=10^{12}$\,cm$^{-2}$ as the
minimum theoretical column density to consider a species
detectable. Below this value we do not take into account the results
of the models.

\subsection{Influence of cosmic rays and UV fields on molecules}

Our modelling results show that most molecular species are sensitive to the presence of external
fields. We checked the variations of their column densities as a
function of UV and CRs strengths (i.e. for the four different
models). A summary of the findings is shown in
Table~\ref{resumen}. While the column density of HOC$^+$ is enhanced
by the presence of UV radiation, many other species are easily
dissociated by UV fields in the external layers of the molecular
clouds (such as CO, HCN, or NS).  On the contrary, we found that
methanol (CH$_3$OH) is clearly destroyed by cosmic rays, while the
production of several other species is favoured by the presence of CR
photons in the ISM of NGC\,1068 (e.g. SiO, CN, or N$_2$H$^+$). Finally, only C$_2$H
shows similar column densities (within one order of magnitude of
variation) for all the models.

Our results agree well with those of \cite{Bayet08}, who also used the UCL\_CHEM model to find suitable molecular tracers of hot cores under different physical conditions. In general, the trends of chemical abundances with respect to $\zeta$ variations found  by \cite{Bayet08} are similar to what is shown in our Table~\ref{resumen}, with few exceptions, CS being the most contrasted species between both works. For this species  \cite{Bayet08} predicted a decrease of its abundance with increased $\zeta$, while our models predict a slight increase. This may be explained by the different physical parameters used in both works (other $\zeta$ strength, initial and final densities, visual extinction and final temperatures).

On the other hand, the comparison of our models with those of \cite{Bayet09a} -where is explored PDR chemistry in a variety of extragalactic environments- leads to big discrepancies regarding the response of the molecular abundances as a function of $\zeta$ and FUV radiation fields. For example, \cite{Bayet09a} deduced that CN and HCO$^+$ are not affected by the variations of $\zeta$, while our results predict enhancements of those species. Also, \cite{Bayet09a} found that CO, HNC, HCN and H$_2$CO are not influenced by changes of the FUV radiation field, whereas our results indicate that the abundances of these molecules are reduced by at least one order of magnitude when the FUV field increases. These disagreements are mainly due to the fact that \cite{Bayet09a} used the UCL\_PDR model where -unlikely the UCL\_CHEM code used here- the depletion of atoms and molecules on to grains, and the subsequent surface chemistry, is not included. In other words, our present modelling assumes that even the gas affected by UV photons has undergone sufficient processing involving gas-grain interactions.

\begin{table}
\caption{Molecular trends with UV fields and cosmic rays in the CND of NGC\,1068}
\centering
\begin{tabular}[!h]{llcccccccc} 
\hline
\hline

Easily dissociated by UV fields &  CO, CS, HCN, CH$_3$OH, SO,\\
				& HCO, SO, NS, HNC, SO$_2$,		\\
				&  N$_2$H$^+$, SiO, CH$_3$CN, H$_2$CS,	\\
				& H$_2$CO, HC$_3$N, CH$_3$CCH	\\
Enhanced by UV fields		& HOC$^+$, 		\\
Easily dissociated by cosmic rays & CH$_3$OH, HC$_3$N		\\
Enhanced by cosmic rays		& CN, HCN, HCO$^+$, HNC, SO,	\\
				& NS, N$_2$H$^+$, SiO, SO$_2$, HOC$^+$	\\
Insensitive			& C$_2$H \\
\hline
\end{tabular}

\label{resumen}
\end{table}

\subsection{Comparison with the observations}

By considering a multi-component model we succeed in reproducing 
the observed column densities (shown in \cite{Aladro12}) of almost all the detected molecules in our
survey  within one order of magnitude, HOC$^+$ being the only exception. We consider this a good match, as the column densities have relevant uncertainties due to the calibrations and the assumption of LTE conditions. In general, the observations are better reproduced by the models which consider a high cosmic ray ionization rate, but low FUV radiation (Model c), and high values of both parameters (Model d).

However, as expected, the scenario depicted by the models reflects clear differences in the origin, as well as timescales of the species, as well as highlighting some degeneracies; nevertheless sometimes the same degeneracies can highlight the prevalence of a particular energetic process. For example CO, ubiquitously tracing gas at a wide range of densities, is well matched by all the models, yet models with high and low radiation field, as long as the cosmic ray ionization rate is high, match the observations better. On the other hand, methanol is much better matched by a model where both radiation field and cosmic ray ionization rates are low: we interpret this `mismatch' in physical conditions between CO and CH$_3$OH as an indication that the two species, on average, arise from different gas; methanol clearly seems to arise from regions where the cosmic ray ionization rate is close to standard, probably not representative of the average cosmic ray ionization rate of NGC\,1068s. This rate is in fact well known to be variable within our own galaxy \cite{Dalgarno06}, and it may well vary in NGC1068 too. On the other hand, the HCO$^+$ observed abundance is only matched if the cosmic ray ionization rate is high (i.e. Models c and d), independently of the FUV field strength.

Unlike for cosmic rays, there are very few species that help us determining the average UV field(s) strength in NGC1068; of particular interest we have HCO, which seems to get closer to the observed value only in models where the UV field is low, while there is no molecule that {\it needs} a high radiation field in order to match the observations. While this does not exclude a high radiation field for this galaxy, especially near the nucleus, it seems unlikely that a rich chemistry comprising of species such as HCO would survive is the average radiation field were to by much higher than the canonical interstellar one.s

Finally it is interesting to note that our models show different recycling cycles among molecules. This behaviour seems to be related to the nature of the species rather than a dependence on the radiation field and cosmic rays strength. For example, the observations of some species, such as CH$_3$OH and NS, are better matched at early times ($\sim10^3$\,yr), since their abundances quickly drop afterwards by at least three orders of magnitude. Similarly, the column densities of some important undetected species (e.g. CH$_3$CCH) drop dramatically below the detectable levels at later times (i.e. $\sim10^5$\,yr). On the other hand, CO, HCO$^+$ and HOC$^+$ always maintain constant values.

%
%
%

\end{document}